\documentclass[aps,prapplied,twocolumn,superscriptaddress,longbibliography]{revtex4-2}
\usepackage{amssymb,amsmath,amsfonts,bm}
\usepackage{epsfig,graphicx}
\usepackage{graphicx}
\usepackage{times}
\usepackage{float}
\usepackage{lipsum}
\usepackage{xcolor}
\usepackage{multirow}
\usepackage{graphicx}
\usepackage{array}

\newcolumntype{P}[1]{>{\centering\arraybackslash}p{#1}}

\usepackage{hyperref}
\hypersetup{colorlinks=true, linkcolor=blue, citecolor=red, urlcolor=blue  }
\usepackage{physics}

\begin{document}
    \title{Engineered Robustness for Nonadiabatic Geometric Quantum Gates}
         
    \author{Xuan Zhang}
    \thanks{These authors contributed equally to this work.}
    \affiliation{Department of Physics and Guangdong Basic Research Center of Excellence for Quantum Science, Southern University of Science and Technology (SUSTech), Shenzhen 518055, China}

    \author{Xiao-Le Li}
    \thanks{These authors contributed equally to this work.}
    \affiliation{Department of Physics, Harbin Institute of Technology, Harbin 150001, China}

    \author{Jingjing Niu}
    \affiliation{International Quantum Academy, Shenzhen 518048, China}
    
    \author{Tongxing Yan}
    \email{yantx@iqasz.cn}
    \affiliation{International Quantum Academy, Shenzhen 518048, China}

    \author{Yuanzhen Chen}
    \email{chenyz@sustech.edu.cn}
    \affiliation{Department of Physics and Guangdong Basic Research Center of Excellence for Quantum Science, Southern University of Science and Technology (SUSTech), Shenzhen 518055, China}

    \date{\today}
    
    \begin{abstract}
While geometric quantum gates are often theorized to possess intrinsic resilience to control errors by exploiting the global properties of evolution paths, this promise has not consistently translated into practical robustness. We present a streamlined framework for nonadiabatic geometric quantum gates (NGQGs) that incorporates additional auxiliary constraints to suppress dynamical contamination and achieve super-robust performance. Within this framework, we also design NGQGs using noncyclic paths, offering enhanced design flexibility. Implemented on superconducting transmon qubits, our scheme realizes high-fidelity single-qubit gates that are robust against Rabi amplitude error $\epsilon$, with infidelity scaling as $\mathcal{O}(\epsilon^4)$, in contrast to the $\mathcal{O}(\epsilon^2)$ behavior of conventional dynamical gates. We further analyze two-qubit NGQGs under parametric driving. Our results identify subtle limitations that compromise performance in two-qubit scenarios, underscoring the importance of phase compensation and waveform calibration. The demonstrated simplicity and generality of our super-robust NGQG scheme make it applicable across diverse quantum platforms.
    \end{abstract}
 
    \maketitle
 
\section{Introduction}    

The geometric phase characterizes how physical quantities twist as they evolve in curved spaces. It manifests itself in numerous seemingly disparate phenomena across disciplines: the midair reorientation of falling cats, the precession of Foucault pendulums, the Aharonov–Bohm effect\cite{aharonov1959significance}, the electronic properties of topological materials\cite{sczhang2011topological}, and the mathematical framework of gauge theories\cite{wilczek1984appearance, simon1983holonomy}. Although the geometric phase in simple quantum systems had been observed in various circumstances\cite{fock1928beziehung, pancharatnam1956phenomenological, aharonov1959significance, aharonov1967observability} prior to its formal understanding, its geometric and global nature remained unrecognized until Berry’s seminal work in 1984\cite{berry1984quantal}. Berry showed that this phase fundamentally arises from cyclic adiabatic evolution in nondegenerate quantum systems. Subsequent generalizations extended the concept to nonadiabatic\cite{aharonov1987phase}, non-Abelian\cite{anandan1988non}, and noncyclic regimes\cite{samuel1988general}, all unified within the mathematical formalism of fiber bundles.

The advent of quantum information science has reinvigorated investigations into geometric phases, particularly through their application in designing fault-tolerant quantum logic gates\cite{zanardi1999holonomic, pachos1999non, ekert2000geometric,  farhi2000quantum, pachos2001quantum, sjoqvist2012non}. Theoretical analyses suggest that geometric quantum gates may outperform conventional dynamical gates in noise resilience, owing to their inherent dependence on global evolutionary trajectories that theoretically confer immunity to local perturbations\cite{ekert2000geometric, zhu2003unconventional, zhu2005geometric,ngqcThomas2011, sjoqvist2012non,  ngqcChen2018, Liu2019, Chen2020, ngqcZhang2020}. However, the presumed robustness of geometric gates remains contentious\cite{blais2003effect, nazir2002decoherence, Kestner2022}. Although Berry phase-based adiabatic geometric gates demonstrate resistance to dephasing and control parameter fluctuations in the adiabatic limit\cite{carollo2003geometric, chiara2003berry, leek2007observation, berger2013exploring, filipp2009experimental}, their operational fidelity is limited by prolonged exposure to decoherence channels\cite{ngqcXiangBin2001}. Conversely, nonadiabatic geometric quantum gates (NGQGs) achieve speeds comparable to those of dynamical gates, yet require meticulous elimination of dynamical phases during nonadiabatic evolution\cite{ngqcXiangBin2001, ngqcZhu2002, sjoqvist2012non, feng2013experimental,ngqcZhao2017, Hong2018a, ngqcChen2018, Xu2018, Liu2019, Xu2020fop, Xu2020prl, Han2020, Liu2021, ding2021nonadiabatic, ding2021path, li2021}. Crucially, experimental robustness proves highly implementation-dependent: geometric gates implementing equivalent logical operations through divergent evolutionary paths may exhibit error susceptibility variations spanning orders of magnitude\cite{ngqcZhu2002, Xu2018, Xu2020fop, li2021, zhou2021nonadiabatic, fang2024nonadiabatic}. Addressing this path-dependent vulnerability constitutes a critical step towards practical geometric quantum gates.

Recent advances propose that additional constraints are essential to realize robust nonadiabatic geometric and holonomic quantum gates\cite{Liu2021}. Experimental verification in superconducting qubits demonstrates that such constraints enable superrobust holonomic gates\cite{li2021}, though their reliance on states in non-computational subspace introduces complications like leakage errors and enhanced decoherence. Building upon these insights, here we apply the general discussion in Ref.\cite{Liu2021} to a simple two-level system, establishing a streamlined framework for NGQGs. Our analysis reveals that controlled implementation of auxiliary constraints systematically enhances gate resilience against control parameter fluctuations. Furthermore, we extend the idea to the construction of two-qubit NGQGs using parametric driving techniques\cite{ bertet2006parametric, strand2013paradrive, caldwell2018parametrically, reagor2018paradrive, li2018paradrive, petrescu2023accurate}. Notably, we identify previously overlooked subtleties that compromise gate robustness in two-qubit NGQGs that rely on parametric driving\cite{ngqcChen2018, chen2020high, Xu2020prl, ding2021nonadiabatic, ding2021path, yang2023experimental, chen2024universal, liang2024nonadiabatic, hong2024unconventional}, highlighting critical considerations for practical implementations.

\section{Single-qubit NGQGs}    

To begin, we briefly review the conventional construction of single-qubit NGQGs. Consider a two-level system subjected to an external drive with real-valued time-dependent amplitude and phase, $\Omega(t)$ and $\phi(t)$. In the rotating frame associated with the free qubit, the Hamiltonian of the system reads ($\hbar=1$):
\begin{equation}\label{hamiltonian}
H_0(t) = \frac{1}{2}[\Omega(t)e^{-i\phi(t)}|0\rangle\!\langle1| + \Omega(t)e^{i\phi(t)}|1\rangle\!\langle0|]
\end{equation}
The evolution of such a driven two-level system can be systematically designed using the Lewis-Riesenfeld dynamical invariant and reverse-engineering methods\cite{Lewis1969dynamicalinvariant, mostafazadeh2001dynamical, chenxi2011dynamicalinvariant, Utkan2012dynamicalinvariant, torrontegui2014dynamicalinvariant}. In the following, we adapt this method but focus on the geometric properties during the evolution. We consider a cyclic process defined by a closed curve in the parameter space of $\Omega(t)$ and $\phi(t)$, with $t \in [0, \tau]$. At each point along the curve, a complete set of auxiliary states $\left|\xi_m(t)\right\rangle$ ($m = 1, 2$) is chosen. Due to the cyclic nature of the process, we have $H_0(0) = H_0(\tau)$ and $\left|\xi_m(0)\right\rangle = \left|\xi_m(\tau)\right\rangle$.

Consider a set of orthogonal states $|\psi_k(t)\rangle$ ($k$ = 1,2) that evolve according to the Schr\"{o}dinger equation, and satisfy the following initial condition: $|\psi_k(0)\rangle=|\xi_k(0)\rangle$. At each moment, $|\psi_k(t)\rangle$ can be related to the above auxiliary states via a unitary transformation: $(|\psi_1(t)\rangle, |\psi_2(t)\rangle)^T = S(t)(|\xi_1(t)\rangle, |\xi_2(t)\rangle)^T$, where $S(t)=\mathcal{T} e^{-i \int_{0}^{t}(A+K)dt'}$ ($\mathcal{T}$ stands for time ordering). $A$ and $K$ are matrices defined by $A_{ij}=i\langle\xi_{i}(t)|\partial_{t}|\xi_{j}(t)\rangle$ and $K_{ij}=-\langle\xi_{i}(t)|H_{0}(t)|\xi_{j}(t)\rangle$. Since $A$ is independent of the Hamiltonian, and depends only on the closed curve in the parameter space and the predefined auxiliary states along the curve, it represents a geometric quantity, whereas $K$ corresponds to a dynamical term. In general, $A$ and $K$ are non-commutative, so the geometric and dynamical parts become mixed. For a cyclic process of a time span $\tau$, it can be easily shown that the evolution operator, defined via $(|\psi_1(\tau)\rangle, |\psi_2(\tau)\rangle)^T = U_0(\tau)(|\psi_1(0)\rangle, |\psi_2(0)\rangle)^T$, is simply $U_0(\tau)=S(\tau)$.

To proceed, we choose the following set of parameterized auxiliary states: $\left|\xi_{1}(t)\right\rangle=\cos(\alpha(t)/2)|0\rangle+\sin(\alpha(t)/2)e^{i\lambda(t)}|1\rangle$ and $\left|\xi_{2}(t)\right\rangle=\sin(\alpha(t)/2)e^{-i\lambda(t)}|0\rangle-\cos(\alpha(t)/2)|1\rangle$, and impose the following conditions to link the two parameters $\alpha$ and $\lambda$ to the amplitude and phase ($\Omega$ and $\phi$) of the drive signal discussed above:
\begin{equation}\label{constraint1}
\begin{aligned}
&\Omega(t)=\frac{\dot{\alpha}(t)}{\sin(\phi(t)-\lambda(t))}, \\ 
&\phi(t)=\lambda(t)-\arctan(\frac{\dot{\alpha}(t)}{\dot{\lambda}(t)\tan(\alpha(t))}).
\end{aligned}
\end{equation}
It can be shown that under such conditions, the off-diagonal elements of $A$ and $K$ cancel each other, and one obtains $A+K=\sigma_z(\frac{1}{2} \dot\lambda \sin\alpha \tan\alpha-\dot\lambda(\sin(\frac{\alpha}{2}))^2)$ (in the basis of $|\xi_m(t)\rangle$; $\sigma_z$ is the Pauli matrix). Here $-\dot{\lambda}(\sin{\frac{\alpha}{2}})^2$ and $\frac{1}{2}\dot{\lambda}\sin{\alpha}\tan{\alpha}$ are the diagonal elements of $A$ and $K$, respectively. Under such conditions, $S(t)=\mathcal{T} e^{-i \int_{0}^{t}(A+K)dt'}$ can be evaluated in a straightforward way and one finds that the evolving states are proportional to the auxiliary states: $|\psi_1(t)\rangle = e^{-i\gamma(t)}|\xi_1(t)\rangle$ and $|\psi_2(t)\rangle = e^{-i\gamma(t)}|\xi_2(t)\rangle$, where $\gamma(t)=\int_0^t(\frac{1}{2} \dot\lambda \sin\alpha \tan\alpha-\dot\lambda(\sin(\frac{\alpha}{2}))^2)dt'$. 

In order to realize a geometric phase, we further impose the following constraint:
\begin{equation}\label{constraint2}
\begin{split}
K_{mm}(\tau)=\frac{1}{2}\int_{0}^{\tau}\dot{\lambda}\sin\alpha \tan\alpha dt=0 \qquad   (m=1,2).
\end{split}
\end{equation}
This condition eliminates the dynamical term accumulated during the evolution, and results in a time evolution operator that consists a pure geometric term $\varphi=-\int_0^{\tau}\dot\lambda(\sin(\frac{\alpha}{2}))^2 dt$ as
$U_{0}(\tau)=e^{i\varphi}|\xi_1(0)\rangle \langle\xi_1(0)|+e^{-i\varphi}|\xi_2(0)\rangle \langle\xi_2(0)|$. When transformed back to the computational basis \{$|0\rangle,|1\rangle$\}, the above $U_{0}(\tau)$ represents a rotation by $2\varphi$ around the axis of $(\pi-\alpha(0), \lambda(0)+\pi)$. To realize an arbitrary single-qubit quantum gate, one simply choose proper initial values for $\alpha$ and $\lambda$, and find proper $\alpha(t)$ and $\lambda(t)$ that satisfy Eq.(\ref{constraint2}) and deliver the desired rotation angle $\varphi$, then impose Eq.(\ref{constraint1}) to obtain the corresponding Hamiltonian. Furthermore, by replacing $|0\rangle$ and $|1\rangle$ with the two basis states of $|01\rangle$ and $|10\rangle$ of a two-qubit system, one can also realize two-qubit gates of an iSWAP type. 

The above analysis underpins gate construction for geometric quantum computation. The key ingredients are a set of properly chosen auxiliary states and the global constraint in Eq. (\ref{constraint2}). The auxiliary states make $A+K$ diagonal, and the constraint removes the residual dynamical phases in the diagonal of $K$. Under these conditions, the implemented gates are purely geometric.

Both geometric and holonomic quantum gates have been conjectured to exhibit robustness against certain types of noise, such as fluctuations in control parameters and environment-induced decoherence\cite{ekert2000geometric, zhu2003unconventional, zhu2005geometric,ngqcThomas2011, sjoqvist2012non,  ngqcChen2018, Liu2019, Chen2020, ngqcZhang2020, carollo2003geometric, chiara2003berry, leek2007observation, berger2013exploring, filipp2009experimental}. Such robustness is believed to stem from the geometric entity in these gates being a global feature determined by the evolution path, rendering it inherently resilient to local perturbations. However, the robustness of geometric gates remains a topic of considerable debate\cite{nazir2002decoherence, blais2003effect, tong2023geometric, Kestner2022}. In the following, we show that the geometric gates constructed above are not inherently robust against fluctuations in the driving pulse amplitude. For other types of error, such as frequency detuning, a similar argument can be formulated.  

To examine this issue, we introduce a perturbation of the form: $\Delta H = V(t)|0\rangle\langle1| + \text{H.c.}$ Comparing $\Delta H$ to the original Hamiltonian $H_0$, it is evident that $\Delta H$ represents a fluctuation in the driving amplitude $\Omega(t)$. It can be easily checked that under this perturbation, the off-diagonal elements of $A$ and $K$ generally do not cancel. Consequently, while gates constructed via the standard procedure (without perturbation) are purely geometric, they become contaminated by dynamical terms in the presence of such a perturbation.

Treating $V(t)$ as a small perturbation relative to the amplitude of $\Omega(t)$, the evolution operator for a cyclic process to first order is\cite{Liu2021}:
\begin{equation}\label{evolution2}
U(\tau)=U_{0}(\tau)[1-i\sum^{2}_{m \neq n}D_{mn}|\xi_{m}(0)\rangle\langle\xi_{n}(0)|]-\mathcal{O}(V^2)
\end{equation}
Here, $D_{mn} = \int_{0}^{\tau} \langle\psi_{m}(t)|V(t)|\psi_{n}(t)\rangle dt$ represents the first-order extra dynamical term induced by $V(t)$. Using this expression, the gate fidelity $F$ is given by:
\begin{equation}\label{fidelity}
F=1-\frac{1}{4}\sum^{2}_{m,n=1}|D_{mn}|^2-\mathcal{O}(V^4).
\end{equation}
Clearly, the extra dynamical term $D_{mn}$ resulting from the control error reduces the fidelity $F$. This demonstrates that achieving NGQGs robust against driving amplitude fluctuations (to at least second order) requires an additional constraint: 
\begin{equation}\label{constraint3}
D_{mn} = \int_{0}^{\tau} \langle\psi_{m}(t)|V(t)|\psi_{n}(t)\rangle dt = 0
\end{equation}
Crucially, previous NGQGs implementations only explicitly imposed the constraint of Eq. (\ref{constraint2}), not $D_{mn}=0$, providing no guarantee for the conjectured robustness. We note that $D_{mn}$ is a path-dependent integral. It is therefore possible that an evolution path constructed without explicitly imposing $D_{mn}=0$ might coincidentally satisfy this condition. In this fortuitous case, the gate operation remains purely geometric, and the theoretically predicted robustness against control errors can be observed. This fact may explain the path-dependent robustness observed in previous NGQGs experiments. The role of $D_{mn}$ in evaluating gate robustness will be further discussed in the context of the experimental results presented in Fig. \ref{robust}.

In the simple case of two-level systems discussed above, satisfying Eq.(\ref{constraint3}) is sufficient for achieving high-fidelity NGQGs. For realistic qubits with multiple levels, such as the transmon type of superconducting qubits, one also needs to consider the cross coupling between the states in and out of the computational space resulted from the perturbation. In either case, such cross coupling represents unwanted dynamical "contamination" and must be eliminated in order to realize high-fidelity geometric quantum gates. 

The above discussion has outlined a general scheme for constructing nonadiabatic, pure geometric single-qubit gates robust against driving amplitude fluctuations. In practice, one can use various optimization methods to search for control pulses satisfying all constraints. A commonly used simple strategy is to employ segmental pulses, with each segment having constant phase parameter $\phi$\cite{ngqcXiangBin2001, ngqcZhu2002, sjoqvist2012non, feng2013experimental,ngqcZhao2017, Hong2018a, ngqcChen2018, Xu2018, Liu2019, Xu2020fop, Xu2020prl, Han2020, Liu2021, ding2021nonadiabatic, ding2021path, li2021}. Such a strategy largely simplifies the analysis and implementation of NGQGs. 

\section{Robust single-qubit gates using open paths}    
In this section, we further demonstrate that robust single-qubit gates can also be readily constructed within the framework established in the previous section, even utilizing noncyclic paths. These gates maintain robustness against driving amplitude fluctuations. Crucially, reducing the restriction of using cyclic paths affords significantly greater flexibility in designing gate pulses.

\begin{figure}
	\centering
	\includegraphics{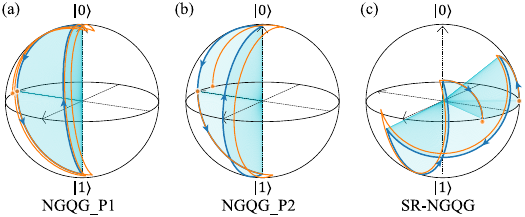}
        \caption{
        The trajectories of the auxiliary state $|\xi_1(t)\rangle$ on the Bloch sphere for an $X$ gate realized by three different NGQG schemes: (a) NGQG\_P1, (b) NGQG\_P2, and (c) SR-NGQG. NGQG\_P1 and NGQG\_P2 are adopted from \cite{ngqcChen2018}. The blue and yellow curves correspond to Rabi errors of $\epsilon=0$ and 0.1, respectively. Notice that the SR-NGQG scheme uses an open path. The NQQGs in (a) and (c) exhibit robustness against Rabi error, as indicated by the relatively small change in their trajectories under error perturbation. In contrast, the NGQG in (b) is not robust, showing a pronounced trajectory shift in the presence of Rabi error. A detailed comparison of their robustness performance is given in Fig.\ref{robust}.
        } \label{fig1}
\end{figure}

Table \ref{tab:noncyclic-pulse} in the Appendix details the pulse parameters for an example evolution corresponding to a noncyclic path in the parameter space (hereinafter referred to as SR-NGQG scheme). Indeed, according to Table \ref{tab:noncyclic-pulse}, for the $X$ gate realized by the SR-NGQG scheme, the initial and final values of ($\alpha,\lambda$) are $(\alpha(0),-\frac{\pi}{6})$ and $(\alpha(0)+3\pi,\frac{7\pi}{6})$. Moreover, it can be shown that the two auxiliary bases $\left|\xi_{1}(t)\right\rangle$ and $\left|\xi_{2}(t)\right\rangle$ do not return to their initial states at the end of evolution (Fig.\ref{fig1}(c) shows the case of $\left|\xi_{1}(t)\right\rangle$). Nevertheless, most analyses from the preceding section remain directly applicable. For example, this pulse satisfies the three constraints specified in Eq. (\ref{constraint1}), Eq. (\ref{constraint2}), and Eq. (\ref{constraint3}). Crucially, satisfying the first two constraints ensures cancellation of the off-diagonal elements of the $A$ and $K$ matrices and sets the diagonal elements of $K$ to zero. This simplification reduces the calculation of the evolution operator to a straightforward integral on the relevant path. It can be easily verified that the overall operator is $-i\sigma_x$, which is equivalent to an $X$ gate. Most importantly, the evolution operator retains its robustness against driving amplitude fluctuations. Below, we present experimental results from a superconducting quantum circuit implementing this gate.

\begin{figure}
	\centering
	\includegraphics{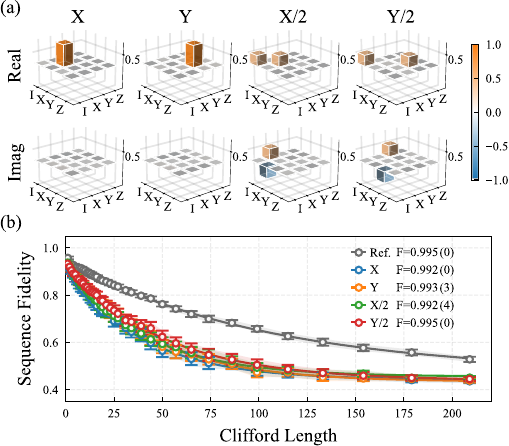}
	\caption{QPT and RB characterization of single-qubit gates realized by our SR-NGQG scheme. (a) Bar charts show the real and imaginary parts of the reduced quantum process matrices $\chi$ for four gates: $X$, $Y$, $X/2$, and $Y/2$, respectively. The solid black outlines are for ideal gates. QPT fidelity is calculated using the experimental data: 0.996($X$), 0.994($Y$), 0.991($X/2$), and 0.993($Y/2$). (b) Sequence fidelity as a function of the number of Clifford gates for both the reference and interleaved RB experiments. Each data point is averaged over 50 random sequences, with the standard deviations plotted as error bars. Fitting the reference curve gives an average gate fidelity of 0.9981 for the single-qubit gates. Fidelity of the four specific gates, $X$, $Y$, $X/2$, and $Y/2$ can be extracted from the difference between the reference and the interleaved curves.} \label{qpt_RB}
\end{figure}

Measurements were performed on transmon-type superconducting qubits (device details can be found in Ref.\cite{Xu2020prl2}). We characterize single-qubit gate performance using quantum process tomography (QPT, Fig.\ref{qpt_RB} (a)) and randomized benchmarking (RB, Fig.\ref{qpt_RB} (b)). The QPT fidelity for the four tested gates is: 0.996($X$), 0.994($Y$), 0.991($X/2$), and 0.993($Y/2$). Reference RB measurements yield an average gate fidelity of 0.995. The interleaved RB analysis gives fidelities of 0.992 ($X$), 0.993 ($Y$), 0.992 ($X/2$), and 0.995 ($Y/2$).

\begin{figure}
	\centering
	\includegraphics{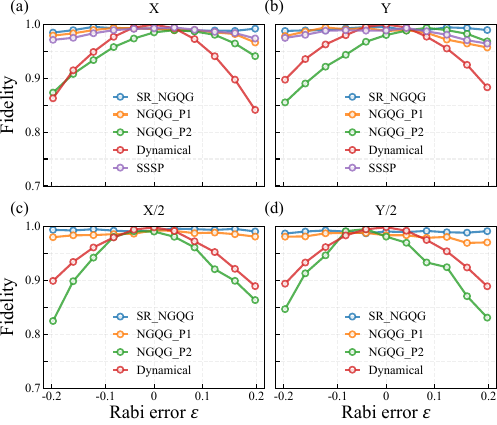}
	\caption{QPT Fidelity as a function of Rabi error for four single-qubit gates. (a)\&(b) $X$ and $Y$ gates implemented using our SR-NGQG scheme, NGQG\_P1, NGQG\_P2, SSSP pulse, and a dynamical gate of a Gaussian pulse. (c)\&(d) $X/2$ and $Y/2$ gates realized using SR-NGQG, NGQG\_P1, NGQG\_P2, and a dynamical gate of a Gaussian pulse.
    }\label{robust} 
\end{figure}

Next, we demonstrate the enhanced robustness against driving amplitude fluctuations for the single-qubit gates. Specifically, we consider a quasi-static (nearly time-independent during gate operation) Rabi error $\varepsilon$ that introduces a perturbation to the Hamiltonian as $\Delta H=\varepsilon\Omega(t)|0\rangle\langle1| + \text{H.c.}$ Figure \ref{robust} compares the QPT fidelity as a function of $\varepsilon$ for different implementations of the following gates: $X$, $Y$, $X/2$, and $Y/2$, including our scheme, NGQGs adopted from \cite{ngqcChen2018}, single-shot-shaped pulse from \cite{VanDamme2017}, and conventional dynamical gates using a Gaussian type of pulse. Detailed information of the pulses used can be found in Table \ref{tab:pulse_robustness} and \ref{tab:noncyclic-pulse} in the Appendix.

In all four gates tested, our scheme consistently demonstrates high robustness across the entire error range. Indeed, as derived from Eq. (\ref{fidelity}), the gate fidelity of our scheme follows $F = 1-\mathcal{O}(\varepsilon^4)$, which is responsible for its superior robustness. The two adopted NGQG pulses show divergent performance: NGQG\_P1 exhibits strong robustness, whereas NGQG\_P2 proves considerably less robust. The single-shot-shaped pulse (SSSP), numerically optimized for robustness, also delivers excellent performance, though at the cost of significantly longer duration. For the dynamical gates, the infidelity can be shown to scale in a straightforward way as $\varepsilon^2$ for all four gates, which aligns reasonably well with the experimental data. Overall, our SR-NGQG scheme achieves an optimal balance between gate efficiency and robustness. Table \ref{tab:robust_integral} in the Appendix compares the value of $D_{12}/\epsilon$ for all the pulses examined in our experiments investigating gate robustness (see Fig.(\ref{robust})). In general, the magnitude of $D_{12}$ serves as a reliable indicator for assessing gate robustness. 

\section{Two-Qubit NGQGs}

Compared with single-qubit gates, two-qubit gates typically exhibit higher error rates due to decoherence, control inaccuracies, and unwanted couplings (e.g., crosstalk and residual interactions). As a result, they constitute the dominant performance bottleneck in quantum circuits, posing major challenges for both quantum error correction and applications in the noisy intermediate-scale quantum (NISQ) era\cite{Oliver2020review}. Developing high-fidelity, noise-resilient two-qubit gates is therefore critical. Geometric and holonomic quantum gates are believed to offer a promising solution: for two-qubit gates confined to a two-level computational subspace -- such as iSWAP -- the schemes developed for single-qubit NGQGs can often be directly extended to realize two-qubit counterparts.

We consider two-qubit gates implementing swap operations within a two-dimensional computational subspace spanned by states ${|\mu_i\rangle}$ $(i=1,2)$ (e.g., $|\mu_1\rangle=|01\rangle$, $|\mu_2\rangle=|10\rangle$ for an iSWAP gate). The Hamiltonian can be written as $H = g_\text{eff}|\mu_1\rangle\langle\mu_2| + \text{H.c.}$ Comparing this with Eq.~(\ref{hamiltonian}), applying the single-qubit NGQGs scheme to the two-qubit case requires the effective coupling $g_\text{eff}$ to possess a tunable amplitude and phase. Such tunability is naturally provided by parametric driving\cite{ bertet2006parametric, strand2013paradrive, caldwell2018parametrically, reagor2018paradrive, li2018paradrive, petrescu2023accurate}, which modulates circuit parameters to activate otherwise inaccessible couplings. In a typical implementation, one qubit's frequency is modulated as $\omega(t) = \omega_0 + A \sin(\Delta t +\phi)$, where $\omega_0$ is the qubit's central frequency, $\Delta = \omega_{\mu_1} - \omega_{\mu_2}$ is the state detuning, and $A$ and $\phi$ specify the driving amplitude and phase. This modulation yields an effective Hamiltonian of the form $ge^{-i\phi}|\mu_1\rangle\langle\mu_2|$ (see Appendix). While this mechanism provides a flexible route for realizing complex-valued tunable couplings, it also introduces additional sensitivity to control fluctuations. As discussed below, this sensitivity can influence the overall robustness of two-qubit NGQGs and thus requires careful calibration and phase management in practical implementations.

Consider a system of two qubits described by the following Hamiltonian where qubit 1 is parametrically driven (lab-frame):
\begin{equation}\label{twoQHamiltonian}
    \begin{aligned}
        H= & (\omega_1^0+A\sin(\Delta t +\phi))a^{\dagger}_1a_1+\omega_2^0a^{\dagger}_2a_2 \\
        & + \frac{\alpha_1}{2} a_1^\dagger a_1^\dagger a_1 a_1 + \frac{\alpha_2}{2} a_2^\dagger a_2^\dagger a_2 a_2 \\
           & +g_{12}(a^{\dagger}_1+a_1)(a^{\dagger}_2+a_2).\\
    \end{aligned}
\end{equation}
Note that realistic realizations of NGQGs schemes often employ segmental pulses (e.g., Table \ref{tab:noncyclic-pulse}). Implementing this approach in the two-qubit system requires segmental parametric driving of the Hamiltonian in Eq. (\ref{twoQHamiltonian}). Specifically, the driving parameters $A$ and $\phi$ may assume different values across segments. For simplicity, we maintain $\phi$ constant within each segment (consistent with Table \ref{tab:noncyclic-pulse}). Furthermore, while Table \ref{tab:noncyclic-pulse} utilizes a time-varying amplitude $\Omega(t)$, this can be replaced by a constant amplitude provided the time-integrated $\int \Omega(t)dt$ per segment (corresponding to the rotation angle of each segment) remains invariant. Consequently, we set $A$ constant within each segment, with values determined by the respective rotation angles. Under these conditions, we define the following rotating frame: 
\begin{equation}\label{twoQrotatingframe}
    \begin{aligned}
      U(t) =& \ e^{-i(\omega_1^0t+\int_0^{t}A\sin(\Delta t^{\prime} +\phi)dt^{\prime})a^{\dagger}_1a_1}e^{-i\omega_2^0ta^{\dagger}_2a_2} \\
      & \times e^{-i\frac{1}{2}\alpha_1 t a_1^\dagger a_1^\dagger a_1 a_1} e^{-i\frac{1}{2}\alpha_2 t a_2^\dagger a_2^\dagger a_2 a_2}\\
           =& \ e^{-i(\omega_1^0t-\frac{A}{\Delta}\cos(\Delta t +\phi)+\phi^{\prime})a^{\dagger}_1a_1}e^{-i\omega_2^0ta^{\dagger}_2a_2} \\
           & \times e^{-i\frac{1}{2}\alpha_1 t a_1^\dagger a_1^\dagger a_1 a_1} e^{-i\frac{1}{2}\alpha_2 t a_2^\dagger a_2^\dagger a_2 a_2}
    \end{aligned}
\end{equation}
Here $\phi^{\prime}$ is an extra phase factor addressing discontinuities in $\phi$ arising from segmental evaluation of the phase accumulation integral $\int_0^{t}A\sin(\Delta t^{\prime} +\phi)dt^{\prime}$. In the rotating frame defined by this transformation, Eq. (\ref{twoQHamiltonian}) becomes (See Appendix for a detailed derivation. Here we only consider the lowest three energy states for each qubit.):  
\begin{equation}\label{twoQHrotating}
    \begin{aligned}
        H^R = & g_{12}\sum_{m=-\infty}^{+\infty}i^m J_m(A/\Delta)e^{im(\Delta t+\phi)}e^{i\phi^{\prime}} \\
        & \left(e^{i\Delta^{10}_{12}t}|10\rangle\langle 01| + \sqrt{2} e^{i\Delta^{11}_{12}t} |11\rangle \langle02| \right.\\
        & +\left. \sqrt{2}e^{i\Delta^{20}_{12}t}|20\rangle\langle 11|\right) +\text{H.c.}
    \end{aligned}
\end{equation}
Here $\Delta^{\nu\mu}_{kl} = \omega_k^0-\omega_l^0 + (\nu-1)\alpha_k -\mu\alpha_l$, where $\alpha_i$ denotes the anharmonicity of qubit $i$, defined as the difference between the energy separation of the first and second excited states and that between the ground and first excited states. $J_m$ represents $m$-th order Bessel function of the first kind.  

By setting the modulating frequency $\Delta$ to be resonant with the energy difference between two specific states $|\mu_1\rangle$ and $|\mu_2\rangle$, a single transition term in Eq. (\ref{twoQHrotating}) (and its conjugate) becomes nonoscillatory. This resonant term dominates the evolution as all oscillating components average to zero, achieving a swap operation between $|\mu_1\rangle$ and $|\mu_2\rangle$ with a phase angle determined by the integration of $H^R$ over the evolution span. For example: When $\Delta = \omega_{|01\rangle} - \omega_{|10\rangle} = \omega_2^0 - \omega_1^0$, the dominant term $ig_{12}J_1(A/\Delta)e^{i\phi}e^{i\phi^{\prime}}|10\rangle\langle 01| + \text{H.c.}$ can be used to implement an iSWAP gate. When $\Delta = \omega_{|02\rangle} - \omega_{|11\rangle} = \omega_2^0 + \alpha_2 - \omega_1^0 = -\Delta_{12}^{11}$, the nonoscillatory term $ig_{12}J_1(A/\Delta)e^{i\phi}e^{i\phi^{\prime}}|11\rangle\langle 02| + \text{H.c.}$ can realize a controlled phase (CPhase) gate (e.g., a specific case is the ubiquitous controlled-Z (CZ) gate in superconducting quantum circuits).

\begin{figure*}
    \centering
    \includegraphics[width=\textwidth]{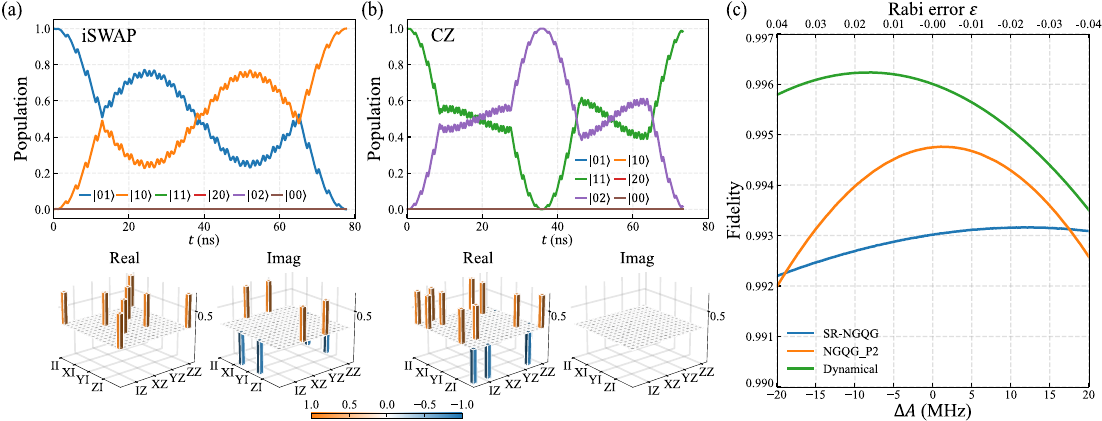}
    \caption{Numerical simulations of two-qubit NGQGs in the rotating frame defined in the main text, using . Panels (a) and (b) show the temporal evolution of state populations (upper) and evolution operators (lower) for iSWAP and CZ gates, respectively. Small oscillations observed in the population is due to off-resonant terms in the Hamiltonian of Eq. (\ref{twoQHrotating}). Gate fidelity is calculated using $\mathrm{Tr}(\tfrac{1}{4}U_\text{exp}U_\text{ideal}^\dagger)$, yielding 99.3\% (iSWAP) and 99.6\% (CZ). (c) Fidelity of three gates as a function of $\Delta A$, fluctuations in the amplitude of parametric driving. The equivalent Rabi error is calculated as $g_{12}J_1(A/\Delta)$.}
    \label{fig:CZ-iSWAP-rot}
\end{figure*}

The single-qubit NGQGs scheme for an $X$ gate (Table \ref{tab:noncyclic-pulse}) can be directly extended to states $|01\rangle$ and $|10\rangle$ to implement a two-qubit nonadiabatic iSWAP gate. However, for CPhase gates, a swap between $|11\rangle$ and $|02\rangle$ introduces a controlled phase in $|11\rangle$, necessitating additional local phase adjustment to achieve a proper CPhase operation. We adopt the CPhase gate sequence from Ref.~\cite{li2021} and single-qubit SR-NGQG scheme for iSWAP, and numerically simulate both two-qubit nonadiabatic iSWAP and CZ gates in the rotating-frame defined by Eq.~(\ref{twoQHrotating}). The simulation results are presented in Fig.~\ref{fig:CZ-iSWAP-rot}(a)\&(b).

We further investigate the robustness of two-qubit NGQGs. As discussed above, for the iSWAP-type NGQG gate, the effective Hamiltonian is $ig_{12}J_1(A/\Delta)e^{i\phi}e^{i\phi^{\prime}}|10\rangle\langle 01| + \text{H.c.}$, indicating an effective Rabi amplitude of $g_{12}J_1(A/\Delta)$. This fact suggests intrinsic robustness against fluctuations in this amplitude. Under the condition where $g_{12}$ and $\Delta$ are fixed, we consider variations in the driving amplitude $A$. We numerically simulate the gate fidelity as a function of $\Delta A$ for three parametrically driven iSWAP gates: a dynamical implementation, the NGQG\_P1 scheme from Ref. \cite{ngqcChen2018}, and the SR-NGQG scheme in this work. The results are shown in Fig. \ref{fig:CZ-iSWAP-rot}(c). 

The most prominent feature is that the maximum fidelity achievable by all three iSWAP gates is inherently limited. As discussed above, in order to realize an iSWAP gate using parametric driving, one must choose proper values for $A$, $\Delta$, and $\phi$ so that only one specific term in Eq.(\ref{twoQHrotating}) dominates. This procedure ignores the contribution of all other terms in Eq.(\ref{twoQHrotating}) and inevitably caps the fidelity. A second key observation is that all three gates exhibit strong robustness across the entire range of $\Delta A$. This can be attributed to the fact that $\Delta A$ influences the effective Rabi amplitude through the first-order Bessel function $J_1(A/\Delta)$, which compresses large deviations in $A$ into small errors in the effective Rabi amplitude. However, it should be noted that in practice the amplitude $A$ can typically be controlled with high precision (on the order of kHz), meaning the range of $\Delta A$ considered here is somewhat exaggerated. Thus, while the observed robustness is notable, it may offer limited practical benefit, whereas the limited maximum fidelity could represent a more significant constraint. 

We would like to add a few comments regarding the experimental implementation of two-qubit NGQGs using parametric driving, although the above discussion seems to suggest that its practical value may be limited. First, the two-qubit state $\Phi(t)$ evolves under the Schrödinger equation $i\hbar \partial\Phi/\partial t=H^R\Phi$ only in the rotating frame defined above. To observe the realized geometric gate, measurements must therefore access $\Phi(t)$. This requires precise knowledge of the relative phase between rotating and laboratory frames for proper phase compensation in measurement pulses (e.g., quantum state tomography). Second, while the above analysis suggests two-qubit NGQGs could be realized by combining single-qubit schemes with parametric driving, significant practical limitations exist beyond the fidelity cap. For example, segmented parametric driving introduces substantial technical complexity: as parametric modulation controls qubit frequency, any pulse fluctuations induce frequency errors that cause spurious phase accumulation, compromising rotating-frame tracking and requiring meticulous calibration for each segment. This is particularly challenging for the transmon qubits, where pulses through finite-bandwidth lines suffer severe distortion, necessitating complex calibration and correction protocols. Finally, the effective Hamiltonian $ig_{12}J_1(A/\Delta)e^{i\phi}e^{i\phi^{\prime}}|\mu_1\rangle\langle \mu_2| + \text{H.c.}$ contains a phase $\phi^{\prime}$ that depends functionally on all the driving parameters of the preceding segments ($A$, $\phi$). Since $\phi$ and $\phi^{\prime}$ collectively determine the rotation axis, fluctuations in $A$ affect both the rotation magnitude (through $J_1(A/\Delta)$) and the orientation of the axis (through the phase composition), which requires careful tracking of phase accumulation, which is technically cumbersome. Overall, if parametric driving is necessary for implementing two-qubit gates, conventional dynamical gates appear more practical.

\section{conclusion}

We have established and experimentally validated a general recipe for constructing nonadiabatic geometric quantum gates that maintain high fidelity in the presence of control imperfections. By imposing an additional constraint to eliminate dynamical contamination, our super-robust protocol achieves consistent suppression of Rabi-error sensitivity, as confirmed by both numerical simulation and experimental results. The scheme is simple, platform-agnostic, and compatible with standard segmented-pulse techniques, making it readily implementable on superconducting qubits and other architectures such as trapped ions, quantum dots, and Rydberg atoms. While the same strategy can, in principle, be extended to two-qubit operations via parametric driving, our analysis highlights hidden sources of fragility that emerge in parametric driving settings, particularly phase discontinuities and waveform distortions. These insights point toward the need for refined control methods and compensation techniques to unlock the full potential of geometric approaches in scalable quantum processors.

The conventional appeal of geometric quantum gates stems from the theoretical notion that the geometric phase is inherently resilient to dynamical noise, as it depends solely on the global properties of the evolution path. However, this perspective has limited practical utility, since real-world noise typically distorts the intended path itself, rather than merely altering the dynamical details along it; consequently, any built-in protection is conditional rather than automatic. Their true value lies in providing a powerful and intuitive design framework. Unlike purely numerical optimization, which may suffer from high-dimensional parameter spaces that hinder efficient convergence to a globally optimal solution, the geometric approach constrains the problem within a physically meaningful structure. By capturing the essential physics of the evolution, it offers a mathematically tractable foundation that significantly reduces the complexity of the search space. This combination of physical intuition and analytical convenience leads to more efficient and elegant solutions, emphasizing that a good framework should not only perform well but also enhance understanding and design efficiency. Thus, while practical robustness must still be actively engineered—often by integrating geometric concepts with modern numerical optimization and error suppression techniques—the geometric framework serves as an invaluable guide for developing high-fidelity, noise-resilient quantum gates.

\section{acknowledgments}
This work was supported by the National Natural Science Foundation of China (No. 12074166), the Guangdong Provincial Key Laboratory (Grant No. 2019B121203002).

\section{Appendix}

\subsection{Pulse sequences}

\begin{table*}
    \centering
    \setlength{\tabcolsep}{12pt}
    \renewcommand{\arraystretch}{1.7} 
    \begin{tabular}{c|c|c|c|c|c}
        \hline\hline
        Gate & $t$ & $\Omega(t)$ & $\phi$ & $\alpha(t)-\alpha(0)$ & $\lambda$ \\
        \hline\hline
        \multirow{4}{*}{$X$} & $(0, \frac{\pi}{\Omega_0})$ & $\Omega_0\sin^2(\Omega_0t)$ & $\frac{\pi}{3}$ & $\frac{1}{2}\Omega_0t-\frac{1}{4}\sin{(2\Omega_0t)}$ & $-\frac{\pi}{6}$ \\
        \cline{2-6}
        & $(\frac{\pi}{\Omega_0}, \frac{3\pi}{\Omega_0})$ & $\Omega_0\sin^2(\frac{\Omega_0t-\pi}{2})$ & $\frac{5\pi}{3}$ & $\frac{1}{2}\Omega_0t-\frac{1}{2}\sin{(\Omega_0t-\pi)}$ & $\frac{7\pi}{6}$ \\
        \cline{2-6}
        & $(\frac{3\pi}{\Omega_0}, \frac{5\pi}{\Omega_0})$ & $\Omega_0\sin^2(\frac{\Omega_0t-3\pi}{2})$ & $\frac{\pi}{3}$ & $\frac{1}{2}\Omega_0t-\frac{1}{2}\sin{(\Omega_0t-3\pi)}$ & $-\frac{\pi}{6}$ \\
        \cline{2-6}
        & $(\frac{5\pi}{\Omega_0}, \frac{6\pi}{\Omega_0})$ & $\Omega_0\sin^2(\Omega_0t-5\pi)$ & $\frac{5\pi}{3}$ & $\frac{1}{2}\Omega_0t-\frac{1}{4}\sin{(2\Omega_0t-10\pi)}$ & $\frac{7\pi}{6}$\\
        \hline
        \multirow{3}{*}{$X/2$} & $(0, \frac{1.28\pi}{\Omega_0})$ & $\Omega_0\sin^2(\frac{\Omega_0t}{1.28})$ & $1.232$ & $\frac{1}{2}\Omega_0t-\frac{1}{4}\sin{(\frac{\Omega_0t}{0.64})}$ & $-0.339$ \\
        \cline{2-6}
        & $(\frac{1.28\pi}{\Omega_0}, \frac{3.28\pi}{\Omega_0})$ & $\Omega_0\sin^2(\frac{\Omega_0(t-1.28\pi)}{2})$ & $-1.236$ & $\frac{1}{2}\Omega_0t-\frac{1}{4}\sin{(\Omega_0t-1.28)}$ & $-2.806$\\
        \cline{2-6}
        & $(\frac{3.28\pi}{\Omega_0}, \frac{4.56\pi}{\Omega_0})$ & $\Omega_0\sin^2(\frac{\Omega_0(t-3.28\pi)}{1.28})$ & $1.232$ & $\frac{1}{2}\Omega_0t-\frac{1}{4}\sin{(\frac{\Omega_0t}{0.64}-3.28)}$ & $-0.339$ \\
        \hline
    \end{tabular}
    \caption{Pulse sequence for a noncyclic robust $X$ and $X/2$ gate (for $Y$ and $Y/2$ gate, simply add $\pi/2$ to $\phi$).} 
    \label{tab:noncyclic-pulse}
\end{table*}

\begin{table*}
    \centering
    \renewcommand{\arraystretch}{1.5} 
    \begin{tabular}{P{2.1cm}|P{0.8cm}P{0.8cm}P{0.8cm}|P{0.8cm}P{0.8cm}P{0.8cm} | P{2.1cm}|P{0.8cm}|P{0.8cm}P{0.8cm}P{0.8cm}P{0.8cm}P{0.8cm}}
    \hline\hline
    \ & \multicolumn{3}{c|}{X gate} & \multicolumn{3}{c|}{X/2 gate} & \multicolumn{7}{c}{SSSP X gate\cite{VanDamme2017}}\\
    \hline\hline
    name & segs & $\theta$ & $\phi$ & seg. & $\theta$ & $\phi$ & $\gamma$ & $\theta$ & $C_1$ & $C_2$ & $C_3$ & $C_4$ & $C_5$ \\
    \hline
    \multirow{5}{*}{{NGQG\_P1}\cite{ngqcChen2018}} 
                              & \multirow{5}{*}{5} & $\pi/2$ & $-\pi/2$ 
                              &                    &         &          
                              & & & & & & &  \\ 
                              &                    & $\pi$   & $3\pi/4$ 
                              & \multirow{3}{*}{3} & $\pi/2$ & $-\pi/2$ 
                              & & & & & & & \\ 
                              &                    & $\pi$   & $-\pi/2$    
                              &                    & $\pi$   & $3\pi/4$ 
    &{\footnotesize{$\chi(0) + 2\alpha + $}} & \multirow{2}{*}{\footnotesize{$5.06\pi$}} & \multirow{2}{*}{\footnotesize{2.3347}} & \multirow{2}{*}{\footnotesize{-1.9450}} & \multirow{2}{*}{\footnotesize{0.3944}} & \multirow{2}{*}{\footnotesize{-0.1139}} & \multirow{2}{*}{\footnotesize{-0.3723}} \\ 
                              &                    & $\pi$   & $3\pi/4$ 
                              &                    & $\pi/2$ & $-\pi/2$ 
    & {\footnotesize{$ \sum_{n} C_n\sin{(2n\alpha)}$}} & & & & & &\\ 
                              &                    & $\pi/2$ & $-\pi/2$ 
                              &                    &         &          
                              & & & & & & & \\ 
    \hline
    \ & \ & \ & \ & \ & \ & \ & $\alpha$ & $\dot{\phi}$ & $\alpha_1$ & $\alpha_2$ & $\alpha_3$ & $\alpha_4$ & $\alpha_5$\\
    \hline
    \multirow{3}{*}{{NGQG\_P2}\cite{ngqcChen2018}} 
                              & \multirow{3}{*}{3} & $\pi/2$ & $-\pi/2$ 
                              & \multirow{3}{*}{3} & $\pi/2$ & $-\pi/2$ 
    &\multirow{2}{*}{{\footnotesize{ $\frac{\pi}{T}t + \sum_{m} \alpha_m$}}} & \multirow{3}{*}{\footnotesize{$\dot{\gamma}\cos{\alpha}$}} & \multirow{3}{*}{\footnotesize{-0.0990}} & \multirow{3}{*}{\footnotesize{-0.1176}} & \multirow{3}{*}{\footnotesize{-0.0394}} & \multirow{3}{*}{\footnotesize{-0.0119}} & \multirow{3}{*}{\footnotesize{0}}\\
                              &                    & $\pi$   & $0$
                              &                    & $\pi$   & $-\pi/4$ 
                              & & & & & & & \\
                              &                    & $\pi/2$ & $-\pi/2$ 
                              &                    & $\pi/2$ & $-\pi/2$
    & {\footnotesize{$\sin{(2m\frac{\pi}{T}t)}$}} & & & & & &\\
    \hline
    \end{tabular}
    \caption{Pulse sequences used in Fig.~\ref{robust}. Left two panels: segmented NGQGs pulses for X and X/2 gates. Here, ``seg.'' denotes the number of segments; $\theta$ is the pulse area ($\int\Omega(t)\,dt$) in each segment, and $\phi$ is the corresponding phase. Within each segment $(\tau_i,\tau_{i+1})$, the envelope is defined as  $\Omega(t)=\Omega_0 \sin^2\!\big[\pi(t-\tau_i)/(\tau_{i+1}-\tau_i)\big]$.
    Right panel: single-shot-shaped pulse (SSSP) for the X gate. In this case, $\alpha$ and $\phi$ describe the trajectory of the parameter states, $\theta$ is the overall pulse area, and $\gamma$ is the global phase. The coefficients $C_n$ and $\alpha_m$ are given up to fifth order in the trigonometric expansion.}
    \label{tab:pulse_robustness}
\end{table*}

\begin{table}
    \centering
    \renewcommand{\arraystretch}{2.0}
    \begin{tabular}{P{2.1cm}|P{2.4cm}|P{2.4cm}}
        \hline\hline
         & $|D_{12}/\epsilon|(X)$ & $|D_{12}/\epsilon|(X/2)$ \\
         \hline\hline
         SR-NGQG & 0.0 & 0.47 \\
         NGQG\_P1 & -0.65 & 0.45 \\
         NGQG\_P2 & 1.57 & 2.67 \\
         Dynamical &-1.57 & -0.78 \\
         SSSP & 0.25 & \\
         \hline
    \end{tabular}
    \caption{Robustness condition integral of different sequences.}
    \label{tab:robust_integral}
\end{table}

The control pulses used in our experiments and simulations are listed in Table~\ref{tab:noncyclic-pulse} (our SR-NGQG scheme) and Table ~\ref{tab:pulse_robustness} (NGQG\_P1, NGQG\_P2, and SSSP schemes). 

\subsection{robustness condition}

The robustness condition given by Eq. (\ref{constraint3}) uses an integral $D_{mn} = \int_0^\tau \langle \psi_m(t)| V|\psi_n(t)\rangle dt$ to characterize the impact of error on the fidelity. Since only Rabi error is considered in this work, the error term $V$ appearing in $D_{mn}$ is proportional to the control Hamiltonian $H_0$ (Eq. (\ref{hamiltonian})). Thus, the integral can be explicitly written as:
\begin{equation}
    D_{mn} = \int_0^\tau \langle \psi_m(t)| \epsilon H_0(t)|\psi_n(t)\rangle dt,
\end{equation}

\noindent where $\epsilon$ is a parameter characterizing the error amplitude and is independent of $H_0(t)$ and the evolving states, therefore it suffices to consider the normalized quantity $D_{mn}/\epsilon$. 

When the cross coupling between states inside and outside the computational subspace is neglected, only two integrals, $D_{12}$ and $D_{21}$, need to be evaluated. Owing to the Hermiticity of the Hamiltonian, $D_{12} = -D_{21}$; thus, either one suffices to quantify the robustness of a given scheme against Rabi error. 

Table~\ref{tab:robust_integral} lists the values of $D_{12}/\epsilon$ for all the pulses used in this work. In general, the magnitude of $D_{12}/\epsilon$ serves as a reliable indicator for assessing gate robustness, as corroborated by Fig.\ref{robust}.

\subsection{Hamiltonian in rotating frame}

We present here the detailed derivation of the rotating-frame transformation for a two-qubit system under parametric driving. The starting point is the lab-frame Hamiltonian given in Eq.~(\ref{twoQHamiltonian}).

In the rotating frame, the effects of single-qubit $Z$ rotations should be removed. Because the frequency of qubit~1 is time dependent, the transformation operator must explicitly include this variation to ensure 
$U^\dagger H_0 U - i\dot{U}U^\dagger = 0$, where $H_0$ denotes the non-interacting part of $H$. This condition is satisfied by choosing
\begin{equation}
    U(t) = \exp\!\left[-i \int_0^t H_0(t^\prime)\, dt^\prime \right],
\end{equation}
since all terms in $H_0(t)$ are diagonal and therefore commute, i.e. $[H_0(t), H_0(t^\prime)]=0$. The explicit form of the transformation operator is then
\begin{equation}\label{Ul2r-qq}
    \begin{aligned}
        U(t) &= \exp\!\Big[-i\big(\omega_1^0 t\, a_1^\dagger a_1 + \tfrac{\alpha_1}{2} a_1^\dagger a_1^\dagger a_1 a_1\big)\Big] \\
             &\quad \times \exp\!\Big[-i\big(\omega_2^0 t\, a_2^\dagger a_2 + \tfrac{\alpha_2}{2} a_2^\dagger a_2^\dagger a_2 a_2\big)\Big] \\
             &\quad \times \exp\!\Big[-i\!\int_0^t A \sin(\Delta t^\prime + \phi)\, dt^\prime \; a_1^\dagger a_1 \Big].
    \end{aligned}
\end{equation}
Equation~\eqref{Ul2r-qq} explicitly separates the free evolution of each qubit from the time-dependent modulation on qubit~1. The first two exponential factors describe the static frequencies $\omega_1^0$, $\omega_2^0$ together with their anharmonicities $\alpha_1$, $\alpha_2$, whereas the last factor incorporates the parametric drive with amplitude $A$, detuning $\Delta$, and phase $\phi$. This unitary defines the rotating frame in which the subsequent interaction Hamiltonian will be derived.

Applying this transformation to the full Hamiltonian yields the rotating-frame interaction Hamiltonian, which captures the effect of parametric driving. 
\begin{equation}\label{Hr-qq}
    \begin{aligned}
        H^{R}(t) &= U^\dagger(t) H U(t) - i \dot{U}(t) U^\dagger(t) \\
                 &= g_{12} \sum_{j,l=1}^{\infty} \sqrt{j(l+1)} \,
                    e^{i \Delta_{12}^{jl} t} e^{-i\Phi(t)} \\
                & \ \ \ \ \ |j,l\rangle\langle j-1, l+1| + \text{H.c.},
    \end{aligned}
\end{equation}
where $\Delta_{mn}^{\nu\mu} = \omega_m - \omega_n + (\nu-1)\alpha_m - \mu\alpha_n$, 
and the phase accumulation from the drive is $ \Phi(t) = \int_0^t A \sin(\Delta t^\prime + \phi)\, dt^\prime. $ For constant driving amplitude $A$ and phase $\phi$, this integral reduces to
\begin{equation}
    \Phi(t) = -\frac{A}{\Delta}\big[\cos(\Delta t + \phi) - \cos\phi\big].
\end{equation}

Applying the Jacobi–Anger expansion,
\begin{equation}
    e^{i\beta \cos(\nu t + \varphi)} = \sum_{n=-\infty}^{\infty} i^n J_n(\beta) e^{in(\nu t + \varphi)},
\end{equation}
the Hamiltonian can be rewritten as
\begin{equation}
    \begin{aligned}
        H^{R} &= g_{12} e^{\tfrac{A}{\Delta}\cos\phi}
        \sum_{j,l=1}^{\infty}\sum_{n=-\infty}^\infty
        i^n J_n\!\left(\tfrac{A}{\Delta}\right) \sqrt{j(l+1)} \\
        &\quad \times e^{i\Delta_{12}^{jl} t} e^{in(\Delta t + \phi)}
        |j,l\rangle\langle j-1, l+1| + \text{H.c.}.
    \end{aligned}
\end{equation}
This expansion makes explicit how the parameter drive generates sideband couplings at harmonics of the modulation frequency. In practice, the Hilbert space is truncated to the lowest three levels, recovering Eq.~(\ref{twoQHrotating}). Whenever the resonance condition $n\Delta = \Delta_{12}^{jl}$ is satisfied, the corresponding term $|j,l\rangle\langle j-1, l+1|$ becomes time independent and mediates a resonant exchange between the two states. For instance, in the iSWAP gate, the states $|10\rangle$ and $|01\rangle$ correspond to $j=1, l=0$, with the resonance condition $\Delta = \omega_1 - \omega_2$. The resulting effective coupling between qubits Q1 and Q2 is
\begin{equation}
    g_{12}^{\text{eff}} = i g_{12} J_1\!\left(\tfrac{A}{\Delta}\right) 
    e^{i(\phi - \tfrac{A}{\Delta}\cos\phi)}.
\end{equation}

In most segmented single-qubit geometric gate schemes, the driving amplitude follows a nonconstant envelope and the driving phase undergoes abrupt jumps between adjacent segments. Consequently, the rotating-frame Hamiltonian is no longer as simple as Eq.~(\ref{Hr-qq}). As discussed in the main text, when extending a single-qubit scheme to two qubits it is common to keep the driving amplitude constant, since the gate operation depends only on the integrated pulse area. In this case, the primary complication arises from phase discontinuities between segments. A sudden change in the control phase introduces a discontinuity in 
$ \Phi(t) = \int_0^t A \sin\!\big(\Delta t^\prime + \phi(t^\prime)\big)\, dt^\prime, $ which in turn alters both the effective coupling strength and its phase. To faithfully reproduce the intended single-qubit control, the effective coupling in the rotating frame must match the ideal parameters for each segment, i.e.$ g^{\mathrm{eff}} = A^{\mathrm{ideal}} e^{i\phi^{\mathrm{ideal}}}. $

Let $A_n$, $\phi_n$, and $\tau_n$ denote the driving amplitude, phase, and duration of the $n$-th segment, respectively. From the single-qubit gate scheme, we obtain the corresponding ideal parameters, denoted $A_n^{\mathrm{ideal}}$ and $\phi_n^{\mathrm{ideal}}$. 

For the first segment, where both amplitude and phase remain constant, the accumulated phase is
\begin{equation}
    \begin{aligned}
        \Phi_1(t) &= \int_0^t A_1 \sin(\Delta t^\prime + \phi_1)\, dt^\prime \\
                  &= -\frac{A_1}{\Delta}\cos(\Delta t + \phi_1) 
                     + \frac{A_1}{\Delta}\cos\phi_1 .
    \end{aligned}
\end{equation}
The corresponding effective coupling strength and phase are
\[
    g_{\mathrm{eff}} = g_{12} J_1\!\left(\tfrac{A_1}{\Delta}\right) 
    e^{\,i\left(\phi_1 - \tfrac{A_1}{\Delta}\cos\phi_1 + \tfrac{\pi}{2}\right)} .
\]
To reproduce the intended single-qubit control, the amplitude and phase in this segment must satisfy
\begin{equation}
    \begin{aligned}
        A_1^{\mathrm{ideal}} &= g_{12} J_1\!\left(\tfrac{A_1}{\Delta}\right), \\
        \phi_1^{\mathrm{ideal}} &= \phi_1 - \tfrac{A_1}{\Delta}\cos\phi_1 + \tfrac{\pi}{2}.
    \end{aligned}
\end{equation}

For the second segment, the accumulated phase becomes
\begin{equation}
    \begin{aligned}
        \Phi_2(t) 
        &= \int_{\tau_1}^t A_2 \sin(\Delta t^\prime + \phi_2)\, dt^\prime + \Phi_1(\tau_1) \\
        &= -\tfrac{A_2}{\Delta}\cos(\Delta t + \phi_2) 
           + \tfrac{A_2}{\Delta}\cos(\Delta\tau_1 + \phi_2) \\
        &\quad - \tfrac{A_1}{\Delta}\cos(\Delta\tau_1 + \phi_1) 
           + \tfrac{A_1}{\Delta}\cos\phi_1 .
    \end{aligned}
\end{equation}
Accordingly, the effective coupling parameters must satisfy
\begin{equation}
    \begin{aligned}
        A_2^{\mathrm{ideal}} &= g_{12} J_1\!\left(\tfrac{A_2}{\Delta}\right), \\
        \phi_2^{\mathrm{ideal}} &= \phi_2 - \tfrac{A_2}{\Delta}\cos(\Delta\tau_1 + \phi_2) \\
                               &\quad + \tfrac{A_1}{\Delta}\cos(\Delta\tau_1 + \phi_1) 
                                 - \tfrac{A_1}{\Delta}\cos\phi_1 + \tfrac{\pi}{2}.
    \end{aligned}
\end{equation}

Abrupt changes in the driving phase lead to the accumulation of additional effective phase. Physically, this occurs because each qubit acquires single-qubit phases during parametric driving. If the rotating frame were continuous, these phases would remain aligned throughout the evolution. However, when the control phase is reset between segments, the rotating frame itself is effectively redefined, and the single-qubit phases must be realigned at the start of each segment. The extra contribution beyond $\phi_n$ therefore serves as a compensating phase. The first two segments illustrate the essential mechanism: phase discontinuities introduce cumulative corrections. Extending this reasoning to the $m$-th segment yields the following compact expressions.
\begin{equation}
    \begin{aligned}
        A_m^{\mathrm{ideal}} &= g_{12} J_1\!\left(\tfrac{A_m}{\Delta}\right), \\
        \phi_m^{\mathrm{ideal}} &= \sum_{k=1}^{m-1} 
        \left[ \tfrac{A_k}{\Delta}\cos(\Delta \tau_k + \phi_k) 
             - \tfrac{A_k}{\Delta}\cos(\Delta \tau_{k-1} + \phi_k) \right] \\
        &\quad + \tfrac{A_m}{\Delta}\cos(\Delta \tau_{m-1} + \phi_m) 
        + \phi_m + \tfrac{\pi}{2}.
    \end{aligned}
\end{equation}

This expression highlights that the effective phase depends not only on the control phase of each segment but also on the corresponding amplitudes. As a consequence, the robustness of geometric gate schemes can be compromised: a design that is robust against Rabi-amplitude errors at the single-qubit level may lose this property when extended to a two-qubit implementation. Hence, the same control scheme does not necessarily exhibit identical robustness in single- and two-qubit configurations.

In summary, although the parametric-driving scheme provides a tunable-phase coupling between two qubits and thus appears well suited for extending single-qubit control schemes to two-qubit systems, segmented control sequences introduce additional phase terms that depend on both the amplitude and phase of each segment. These extra terms complicate the effective coupling and can ultimately compromise the expected robustness of geometric gate schemes when applied to two-qubit operations.


%

\end{document}